\documentclass[runningheads]{llncs}

\usepackage{amssymb}
\usepackage{latexsym}
\usepackage{longtable}

\usepackage{url}

\usepackage{hyperref}

\usepackage{lscape} 
\usepackage{tabulary}
\usepackage{tabularx, rotating}

\usepackage[utf8]{inputenc}
\usepackage[T1]{fontenc}
\usepackage{pifont}
\usepackage{pmboxdraw}
\usepackage[dvipsnames]{xcolor}

\begin{document}

\authorrunning{G. Gonz\'alez et al.}
\titlerunning{UMLS-ChestNet}


\begin{frontmatter}

\title{UMLS-ChestNet: A deep convolutional neural network for radiological findings, differential diagnoses and localizations of COVID-19 in chest x-rays}

\hyphenation{ex-ten-ded}
\hyphenation{suc-cess-ful-ly}
\hyphenation{pneu-mo-nia}
\hyphenation{pro-mi-sing}
\hyphenation{pro-ce-ssing}
\hyphenation{me-tho-do-lo-gy}

\author{
Germán González\inst{1,2} \and 
Aurelia Bustos\inst{3} \and
José María Salinas\inst{4} \and
Maria  de la Iglesia-Vaya\inst{4} \and
Joaquín Galant\inst{5} \and
Carlos Cano-Espinosa\inst{2} \and
Xavier Barber\inst{4} \and
Domingo Orozco-Beltr\'an\inst{6} \and
Miguel Cazorla\inst{2}\and
Antonio Pertusa\inst{2}}


\institute{
Sierra Research S.L, Avda. Costa Blanca 132 EntC. Alicante.  03540, Spain \and 
Universidad de Alicante, Avda. San Vicente, S/N, San Vicente del Raspeig (Alicante), 03690, Spain \and 
Medbravo, C/ Deportista Juan Matos, 4 Alicante, Spain- 03016\and
Unidad Mixta de Imagen Biomédica FISABIO-CIPF. Fundación para el Fomento de la Investigación Sanitario y Biomédica de la Comunidad Valenciana.  \\ Avda. de Catalunya, 21. 46020 Valencia, España\and
Hospital San Juan de Alicante. Carretera de Valencia S/N 03550\\ San Juan de Alicante, Alicante, España \and
Universidad Miguel Hernández. Carretera de Valencia S/N 03550\\ San Juan de Alicante, Alicante, España
}

\maketitle

\begin{abstract}
In this work we present a method for the detection of radiological findings, their location and  differential diagnoses from chest x-rays. Unlike prior works that focus on the detection of few pathologies, we use a hierarchical taxonomy mapped to the Unified Medical Language System (UMLS) terminology to identify 189 radiological findings, 22 differential diagnosis and 122 anatomic locations, including ground glass opacities, infiltrates, consolidations and other radiological findings compatible with COVID-19. We train the system on one large database of 92,594 frontal chest x-rays (AP or PA, standing, supine or decubitus) and a second database of 2,065 frontal images of COVID-19 patients identified by at least one positive Polymerase Chain Reaction (PCR) test. The reference labels are obtained through natural language processing of the radiological reports. On 23,159 test images, the proposed neural network obtains an AUC of 0.94 for the diagnosis of COVID-19. To our knowledge, this work uses the largest chest x-ray dataset of COVID-19 positive cases to date and is the first one to use a hierarchical labeling schema and to provide interpretability  of the results, not only by using network attention methods, but also by indicating the radiological findings that have led to the diagnosis.
\end{abstract}


\end{frontmatter}


\section{Introduction}

The SARS-CoV2 pandemic has led researchers to develop deep learning methods on x-rays (CR/DX) and Computed Tomography (CT) images for the diagnosis of patients with COVID-19. While most methods are focused on the classification of images into the categories COVID-19, other pathologies or no pathology, the early detection and localization of lesions such as infiltrates, in particular ground glass opacities, is essential both for the diagnosis and to predict the evolution of the patient in order to help making clinical decisions.

The typical findings that suggest a COVID-19 infection are ground glass opacities that, even in the initial stages, affect both lungs, particularly the lower lobes, and especially the posterior segments, mainly with a peripheral and subpleural distribution ~\cite{Shi2020,YuenFrankWong2020,Mei-sze2020}. These findings were present in 44\% of patients in the first two days on chest CT scans \cite{Bernheim20}, in 75\% of patients in the first four days \cite{Pan20} and in 86\% of patients during illness days 0-5 \cite{Wang20}.


\begin{figure}[t!]
    \centering
    \begin{tabular}{m{4cm}|l|l}
    \includegraphics[width=4cm, trim={2cm 2cm 2cm 2cm}]{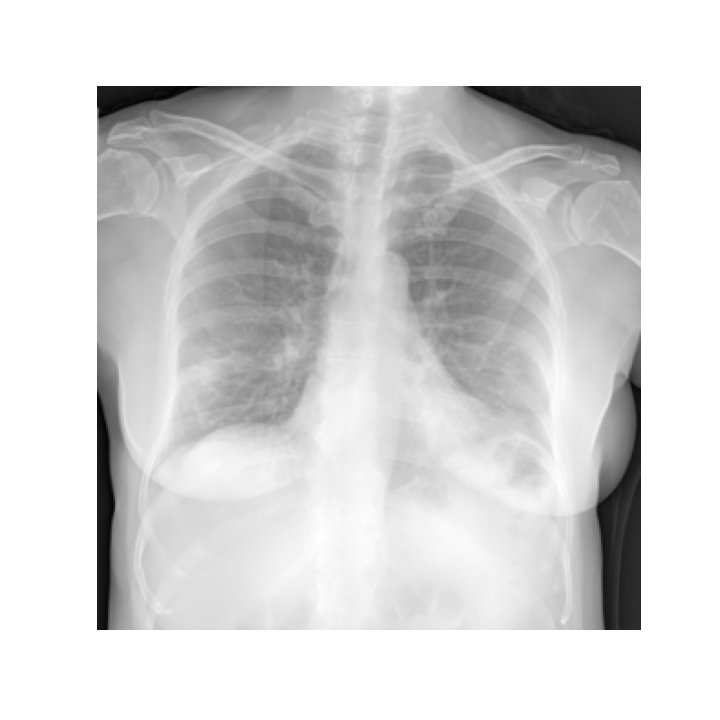} &  
        \tiny
        \begin{tabular}{l}
            
             \underline{Differential diagnosis}\\
            ├── pneumonia \\
            │\verb!   !└── atypical pneumonia\\
            │\verb!   !\verb!   ! └── viral pneumonia\\
            │\verb!   !\verb!   !\verb!   !  ├── \textbf{COVID-19} \\
            ... \\
            \underline{Radiological Findings}  \\
            ... \\
            ├── infiltrates\\
            │\verb!   !├── interstitial pattern\\
            │\verb!   !│\verb!   !├── ground glass pattern \\
            ...\\
            │\verb!   !└── alveolar pattern\\
            │\verb!   !\verb!   ! ├── consolidation\\
            ...\\
            ├── \textbf{increased density}\\
        \end{tabular}             
             
    &
    \tiny
     \begin{tabular}{l}
        
             \underline{Differential diagnosis}\\
            ├── pneumonia \\
            │\verb!   !└── atypical pneumonia\\
            │\verb!   !\verb!   ! └── viral pneumonia\\
            │\verb!   !\verb!   !\verb!   !  ├── \textbf{COVID-19} \\
            ... \\
            \underline{Radiological Findings}  \\
            ... \\
            ├── infiltrates\\
            │\verb!   !├── interstitial pattern\\
            │\verb!   !│\verb!   !├── \textbf{ground glass pattern} \\
            ...\\
            │\verb!   !└── alveolar pattern\\
            │\verb!   !\verb!   ! ├── \textbf{consolidation}\\
            ...\\
            ├── \textbf{increased density}\\
        \end{tabular}

    \\
    \end{tabular}
    
    \caption{Left: Example image of the BIMCV COVID-19 dataset. Middle: nodes of the UMLS hierarchy that are positive, as extracted from the radiology report, are marked in bold. Right: In bold, UMLS nodes positive obtained by the network. Please note that the network detects ground glass patterns and consolidations that are not present on the radiology report. Both radiological findings are highly related to the diagnosis of COVID-19. }
    \label{fig:example}
\end{figure}

Therefore, beyond classifying an image as COVID/No/Others, it is very important to obtain the radiological findings in order to understand the severity of the disease and to predict the possible evolution of the patient. Moreover, it should be taken into account that some PCR positive COVID-19 patients may show a normal RX, therefore only with the image information it may not be possible to perform a correct diagnosis. Similarly, not all radiological findings compatible with COVID-19, such as ground glass opacities, are indeed produced by a Sars-Cov-2 infection. Unlike diagnoses, findings are reported by the radiologists and they can be seen in the images, being less dependent on other clinical variables. 

An example of the proposed methodology can be seen in Fig.~\ref{fig:example}. A chest x-ray (left) is tagged by processing its radiology report with natural language processing to a set of hierarchical labels (middle). The network predicts such labels that include not only the presence of the disease (COVID-19), but also the radiological findings that have lead to it. Please note that the model detects two findings that were not present on the radiology report but are consistent with COVID-19. The proposed method is not constrained to COVID-19, but contains an extense set of labels that encompasses most of the diseases and radiological findings that can be interpreted from a chest x-ray.


Deep learning methods are widely used for chest x-ray pathology detection. In order to train these models, it is essential to have a large dataset of images properly annotated. In the literature there are many datasets intended for this task such as \cite{irvin2019chexpert,johnson2019mimic,wang2017chestx}, with a maximum of 14 output variables associated with them, and none with ground glass opacities or infiltrates as part of their covariates. To the best of our knowledge, the only open dataset currently labeled with a large set of radiological findings is PadChest \cite{bustos2019padchest}. PadChest includes more than 160,000 images obtained from 67,000 patients that were interpreted and reported by radiologists at San Juan Hospital (Spain) from 2009 to 2017, covering six different position views, and including additional information on image acquisition and patient demography. Padchest labels were extracted from the radiological reports, resulting in 22 differential diagnoses, 122 anatomic locations and 189 different radiological findings mapped onto standard Unified Medical Language System (UMLS) using controlled biomedical vocabulary unique identifiers (CUIs) that are organized into semantic hierarchical concept trees. Padchest is the only open dataset that provides labels for consolidations, pneumonia and ground glass opacities, as needed to diagnose COVID-19. 

Different neural network architectures have been proposed in the literature for the classification of pathologies  \cite{rajpurkar2017chexnet,varshni19} from RX images using the datasets from \cite{irvin2019chexpert,johnson2019mimic,wang2017chestx}. However, to the best of our knowledge, there is no previous work addressing the classification of a large set of radiological findings as described by \cite{bustos2019padchest}.


Some recent models based on Convolutional Neural Networks (CNN) have been proposed for the detection of COVID 19. Most of them can be found in recent review articles, such as~\cite{ShiFeng2020,Latif2020,Burlacu2020}. 
However, they all rely on the same public dataset \cite{Cohen2020} for positive samples~\cite{Li2020,Ghoshal2020,Apostolopoulos2020}, or subsequent works derived from it~\cite{Karim2020,Wang2020,Farooq2020,Chowdhury2020}. The dataset from~\cite{Cohen2020} is obtained by gathering images from other publications, which are indeed of low spatial and bitdepth resolution. As negative samples the cited work uses data from the RSNA challenge \cite{RSNA} or from the Kaggle competition~\cite{Kaggle}, which is composed of pediatric images. Moreover, it should be considered that there is a domain mismatch between the positives and the negatives, and therefore, it may happen that models would find events on the images that are not related to the disease but on the domain itself. 




\section{Material and methods}
\label{sec:materialmethods}

\subsection{Dataset}

In this work we use the images and labels from PadChest \cite{bustos2019padchest} and BIMCV COVID-19+ \cite{BIMCVCOVID19}. Both contain labels that are mapped into NLM Unified Medical Language System (UMLS) controlled biomedical vocabulary unique identifiers, referred as CUIs, along with the localizations. Four additional labels with special coding rules were explicitly defined: normal, exclude, unchanged and suboptimal study. Although the excerpts from the report are provided in Spanish, labels are mapped onto CUI codes, thus making the dataset usable regardless of the language. In addition, the labels are structured using a tree structure. 

\begin{table}[]
    \tiny
    \centering
    \begin{tabular}{|l|l|l|}
\hline
        \begin{tabular}{l}
            \textbf{Radiological Findings}  \\
            ├── infiltrates\\
            │\verb!   !├── interstitial pattern\\
            │\verb!   !│\verb!   !├── ground glass pattern\\
            │\verb!   !│\verb!   !├── reticular interstitial pattern\\
            │\verb!   !│\verb!   !├── reticulonodular interstitial pattern\\
            │\verb!   !│\verb!   !└── miliary opacities\\
            │\verb!   !└── alveolar pattern\\
            │\verb!   !\verb!   ! ├── consolidation\\
            │\verb!   !\verb!   ! │\verb!   !└── air bronchogram \\
            │\verb!   !\verb!   ! └── air bronchogram\\
        \end{tabular}
    &         
            \begin{tabular}{l}
            \textbf{Differential diagnosis}\\
            ├── pneumonia \\
            │\verb!   !└── atypical pneumonia\\
            │\verb!   !\verb!   ! └── viral pneumonia\\
            │\verb!   !\verb!   !\verb!   !  ├── COVID-19 \\
            │\verb!   !\verb!   !\verb!   !  └── COVID-19 uncertain\\
            ├── tuberculosis\\
            │\verb!   !└── tuberculosis sequelae\\
            ├── lung metastasis\\
            ├── lymphangitis carcinomatosa\\
            ├── lepidic adenocarcinoma\\
            \end{tabular}
    & 
            \begin{tabular}{l}
            \textbf{Localization}\\ 
            ├── extracorporal\\
            ├── cervical \\
            ├── soft tissue \\
            │\verb!   !├── subcutaneous \\
            │\verb!   !├── axilar \\
            │\verb!   !└── pectoral \\
            │\verb!   !\verb!   ! └── nipple\\
            ├── bone \\
            │\verb!   !├── shoulder \\
            │\verb!   !│\verb!   !├── acromioclavicular \\
            \end{tabular}\\
    \hline
    \end{tabular}
    \caption{Examples of the hierarchical radiological findings, diagnoses and locations present on the reference standard. For the full tree we refer the reader to the supplementary material.}
    \label{tab:UMLS}
\end{table}

Some examples of the tree for differential diagnosis, radiological findings and localizations are shown in Table~\ref{tab:UMLS}. Overall, PadChest contains more than 160,000 images with 189 different radiographic findings, 22 differential diagnoses and 122 anatomic locations. The 92,594 frontal images of the PadChest dataset were used for this work.

Since PadChest's data were collected until 2017, it does not include COVID-19 labels. However, the recently released BIMCV COVID-19+ dataset, whose acquisition methodology and organization is based on PadChest, is intended for training machine learning methods to detect this pathology. This is currently the largest open chest X-ray COVID-19 dataset. It contains chest CXR (CR, DX) and computed tomography (CT) imaging of COVID-19 patients along with their radiographic findings, pathologies, Polymerase Chain Reaction (PCR), Immunoglobulin G (IgG) and Immunoglobulin M (IgM) diagnostic antibody tests, and radiological reports from Medical Imaging Databank in Valencian Region Medical Image Bank (BIMCV).

Like in PadChest, the BIMCV COVID-19+ labels cover a wide spectrum of thoracic entities, images are stored in high resolution and entities are localized with anatomical labels in a Medical Imaging Data Structure (MIDS) format. In addition, 18 image studies were annotated by a team of expert radiologists to include semantic segmentation of radiological findings. Moreover, it provides extensive information, including the patient’s demographic information, type of projection and acquisition parameters for the imaging study, among others. 

The BIMCV COVID-19+ iteration used in the presented work includes 1,380 CR and 885 DX that were added to the full PadChest samples for training, testing and validation. COVID-19 is a pathology where RT-PCR is considered the gold standard for diagnosis of COVID-19 pneumonia~\cite{Kim2020}. Its pattern of radiological findings, even if typical, are not pathognomonic, i.e. their presence does not mean that COVID-19 pneumonia is present beyond any doubt and other pathologies as other atypical and viral pneumonias need to be ruled out. For this reason, COVID-19 pneumonia was included in the taxonomy tree under the differential diagnosis branch \cite{BIMCVCOVID19}.

\subsection{Image preprocesing}

Only frontal images, either with PA or AP, supine, decubitus, lordotic or standing were included in this work. Such images were identified in the databases by inspecting the manual annotations. Images with photometric interpretation of \texttt{MonochromeI}, as signaled on the DICOM fields were inverted. Image intensities were normalized by subtracting the mean dividing by the variance before using them to train any of the proposed networks.

In order to normalize the image size, we first cropped a centered square region of interest, with the size of the smallest image axis. Such technique is preferred to interpolating to a square grid in order to prevent anatomical deformations, and also preferred to image padding to prevent misleading the network at the radiography edges. Images were finally re-scaled to $299\times299$ pixels to be processed by the neural network.

\subsection{Label propagation}

The labels for each image are mapped to the UMLS hierarchy. They are extracted from the NLP interpretation of the radiological reports. Only few nodes of the UMLS tree are extracted from each radiology report. The UMLS hierarchy has an is-a relationship for both the differential diagnoses and radiological findings trees, and an is-part-of relationship for the anatomical location tree. We therefore construct for each image a target vector that places '1' on each node that is in the interpretation of the radiology report or any of their parents until the root. For instance, if the image is associated with a COVID-19 diagnosis as interpreted from the radiological report, and using the tree shown in  Fig.~\ref{tab:UMLS}, we would flag the image as being positive for \textit{viral pneumonia}, \textit{atopical pneumonia}, \textit{pneumonia} and \textit{differential diagnosis}. An image with a label of \textit{pneumonia} would not have a COVID-19 diagnosis.

\subsection{Training}

We evaluate three image classification topologies in this work: EfficientNetB4 \cite{effnet}, Inception-Resnet-V2 \cite{inceptionresnetv2} and XNet \cite{xnet}. Each of these architectures is extended by using two fully connected layers of 512 units with rectified linear units (ReLU) activation and a final classification layer of length 306 with sigmoid activation. Dropout layers with dropout coefficient of $0.2$ were interleaved between the fully connected layers. The model weights were pre-trained using the ImageNet \cite{imagenet} database.

The loss function employed is a binary cross entropy for each of the output dimensions independently, since correlations between dimensions are already handled by the construction of the target labels through the label propagation method.  The networks are trained using adaptive momentum optimization with a learning rate schedule that decreases the learning rate between $10^{-3}$ for the first epochs to $10^{-6}$ for the last ones. Training is performed for $50$ epochs and the model with the best validation accuracy was selected.

The training set consists of 50,000 frontal images from Padchest and 1,000 images from the BIMCV COVID-19+ dataset. 20,000 images from Padchest and 500 from COVID-19 were used for validation. The test set, from which the evaluation results are reported, is composed by 22,594 images from Padchest and 565 images from the BIMCV COVID-19+ dataset. The images are distributed in such manner that two images from the same patient can not appear in the training, validation or test sets.

\subsection{Statistical Analysis}
ROC curves were computed with Scikit-learn. 95\% confidence intervals, when reported, were computed using MedCalc and with R statistical software.

\section{Results}
\label{sec:results}

\begin{table}[t]
\centering
\begin{tabular}{|l|r|r|r|r|}
\hline
                        Variable &      N &  EfficientNetB4 &  InceptionResnetV2 &  XCeption \\
\hline \hline
Avg AUC &  23,159 &        0.787 (0.176) &           0.738 (0.172) &  0.768 (0.175) \\
\hline \hline

                         pacemaker &    367 & 0.997 (0.996-0.999) & 0.997 (0.996-0.998) & 0.997 (0.996-0.999)\\
             single chamber device &     69 & 0.992 (0.990-0.993) & 0.989 (0.987-0.992) & 0.991 (0.990-0.993)\\
                        sternotomy &     382 & 0.998 (0.998-0.999) & 0.969 (0.962-0.976) & 0.996 (0.995-0.997) \\
               dual chamber device &   164 & 0.987 (0.975-0.999) & 0.986 (0.975-0.997) & 0.987 (0.975-0.999) \\
     art. mitral heart valve &     56 & 0.992 (0.987-0.996) & 0.979 (0.971-0.988) & 0.989 (0.983-0.995)\\
                   pulmonary edema &     53 & 0.960 (0.945-0.976) & 0.956 (0.943-0.969) & 0.955 (0.939-0.971) \\
 res. cent. ven. catheter &      66 & 0.962 (0.938-0.987) & 0.907 (0.870-0.945) & 0.965 (0.941-0.989) \\
                        cavitation &    77 & 0.942 (0.916-0.969) & 0.939 (0.918-0.960) & 0.951 (0.931-0.971) \\
                        ...        &     ...&        ...     &             ...     &  ....     \\
                mastectomy &    109 & 0.811 (0.769-0.852) & 0.656 (0.603-0.709) & 0.748 (0.701-0.795)\\
                diaphr. eventration &   182 & 0.779 (0.748-0.809) & 0.717 (0.684-0.750) & 0.718 (0.685-0.751) \\
               callus rib fracture &   414 & 0.760 (0.740-0.781) & 0.711 (0.689-0.733) & 0.742 (0.720-0.764) \\
                    tracheal shift &     148 & 0.763 (0.729-0.797) & 0.658 (0.619-0.697) & 0.790 (0.755-0.824)\\
                      air trapping &    798 & 0.751 (0.734-0.768) & 0.707 (0.690-0.725) & 0.720 (0.703-0.738)\\
               flattened diaphragm &  136 & 0.743 (0.704-0.782) & 0.704 (0.662-0.747) & 0.717 (0.672-0.761) \\
           meds. lipomatosis &     76 & 0.713 (0.659-0.767) & 0.688 (0.634-0.742) & 0.653 (0.592-0.714)\\
                       azygos lobe &     74 & 0.562 (0.504-0.619) & 0.608 (0.548-0.667) & 0.623 (0.566-0.680) \\
\hline
\end{tabular}
\caption{Performance of the three different networks in the top 8 and bottom 8 diagnoses and radiological findings. In parenthesis we show 95\% confidence intervals. The average AUCs are obtained by averaging the AUCs of all predicted variables and are shown with standard deviations in parenthesis.}
\label{tab:several_networks}
\end{table}

\subsection{One class versus all}

First, the performance of the method is evaluated in a per-label setting. As such, we consider as positive for a radiological finding, differential diagnosis or location, any image that has such variable or any of its children on the reference standard interpretation of the radiology report. This allows us to obtain a per variable positive and negative class. The comparison results of the three networks favors EfficientNetB4, with an average AUC over all radiological findings, locations and diagnoses of 0.787, in comparison with XCeption (AUC of 0.768) or InceptionResnetV2 (AUC of 0.737). Table~\ref{tab:several_networks} shows the top eight and bottom eight diagnoses or radiological findings detected by each network.  In  the supplementary material we show the performance for the whole UMLS tree.


\subsection{COVID-19 Results}

\begin{figure}[t!]
  \centering

   \begin{tabular}{cc}
      \includegraphics[trim={0 0 0cm 0}, width=0.45\textwidth]{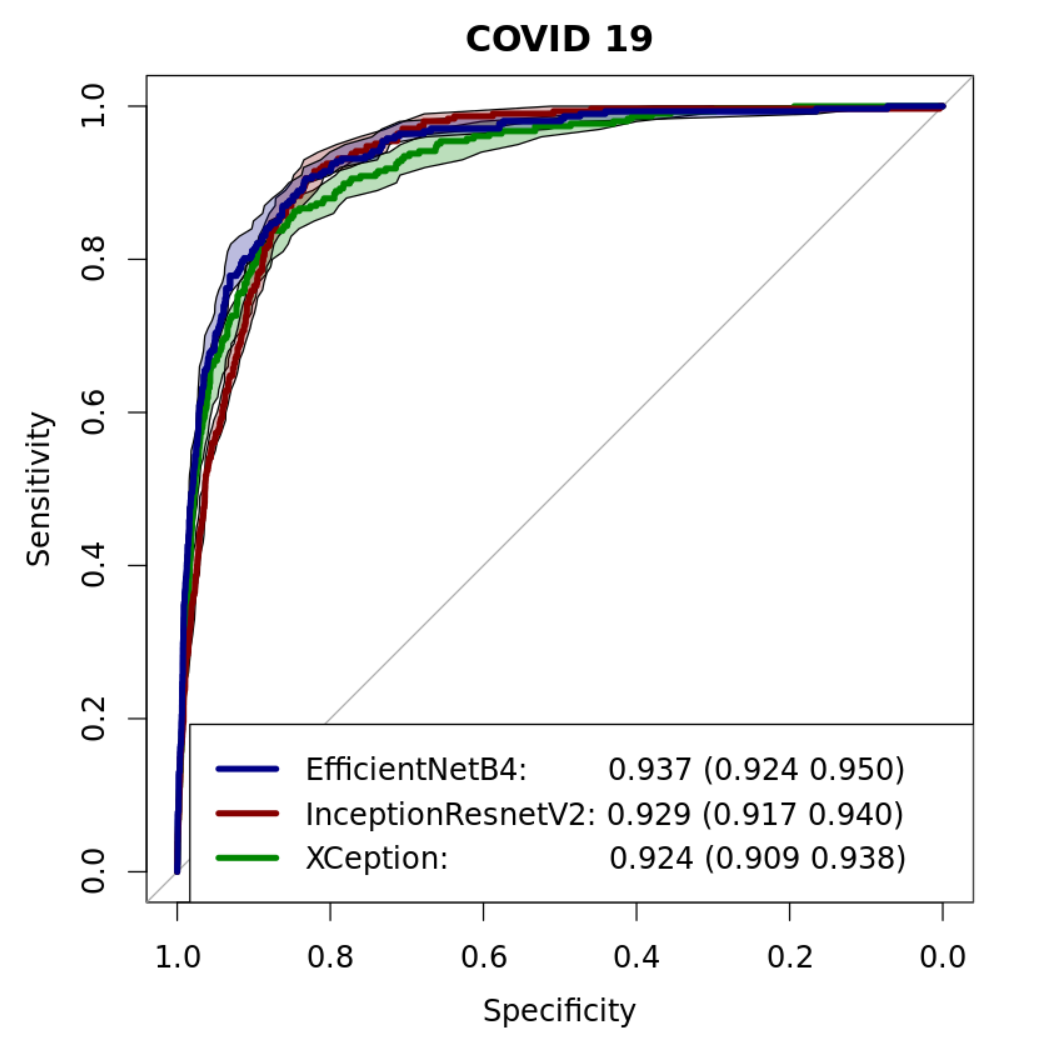} & 
    \includegraphics[trim={0 0 0cm 0}, width=0.45\textwidth]{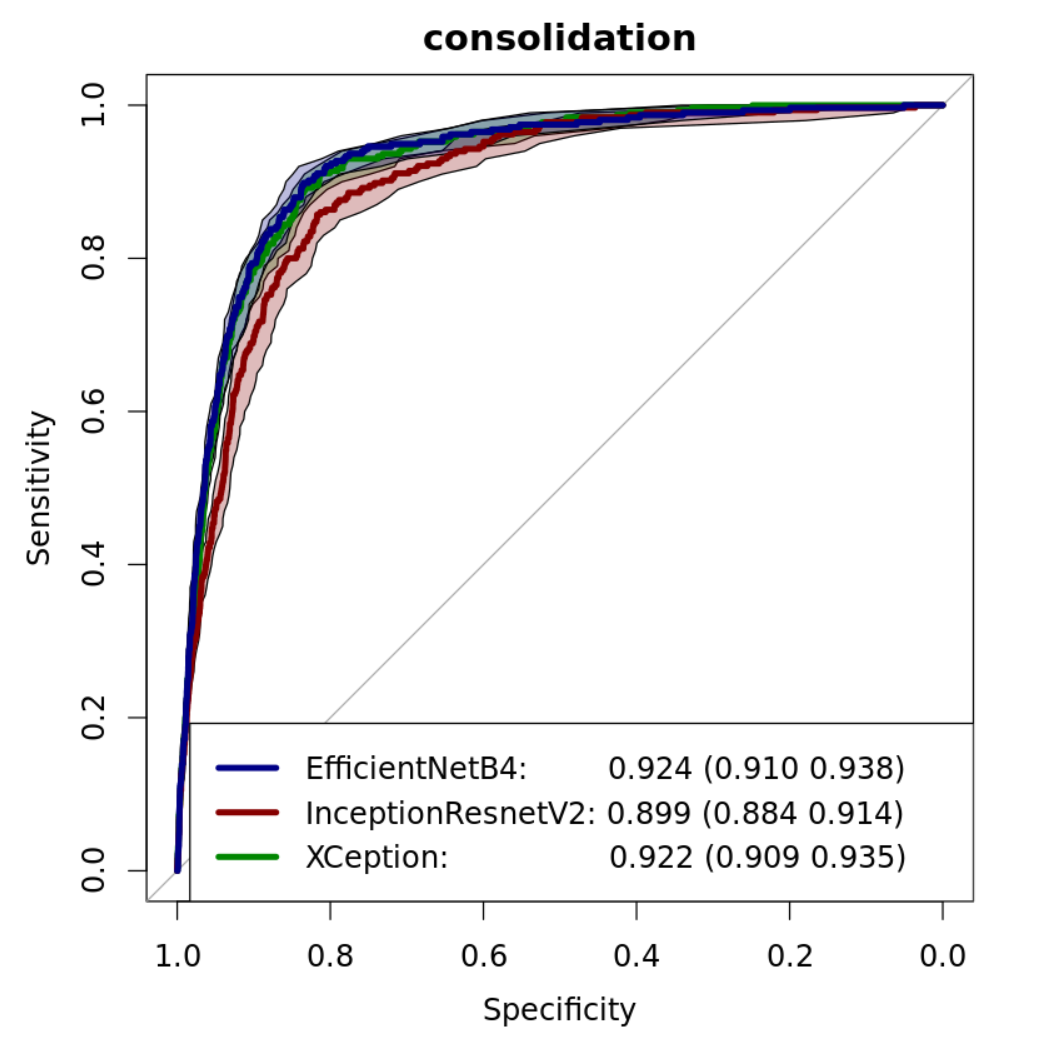} \\
        \includegraphics[trim={0 0 0cm 0}, width=0.45\textwidth]{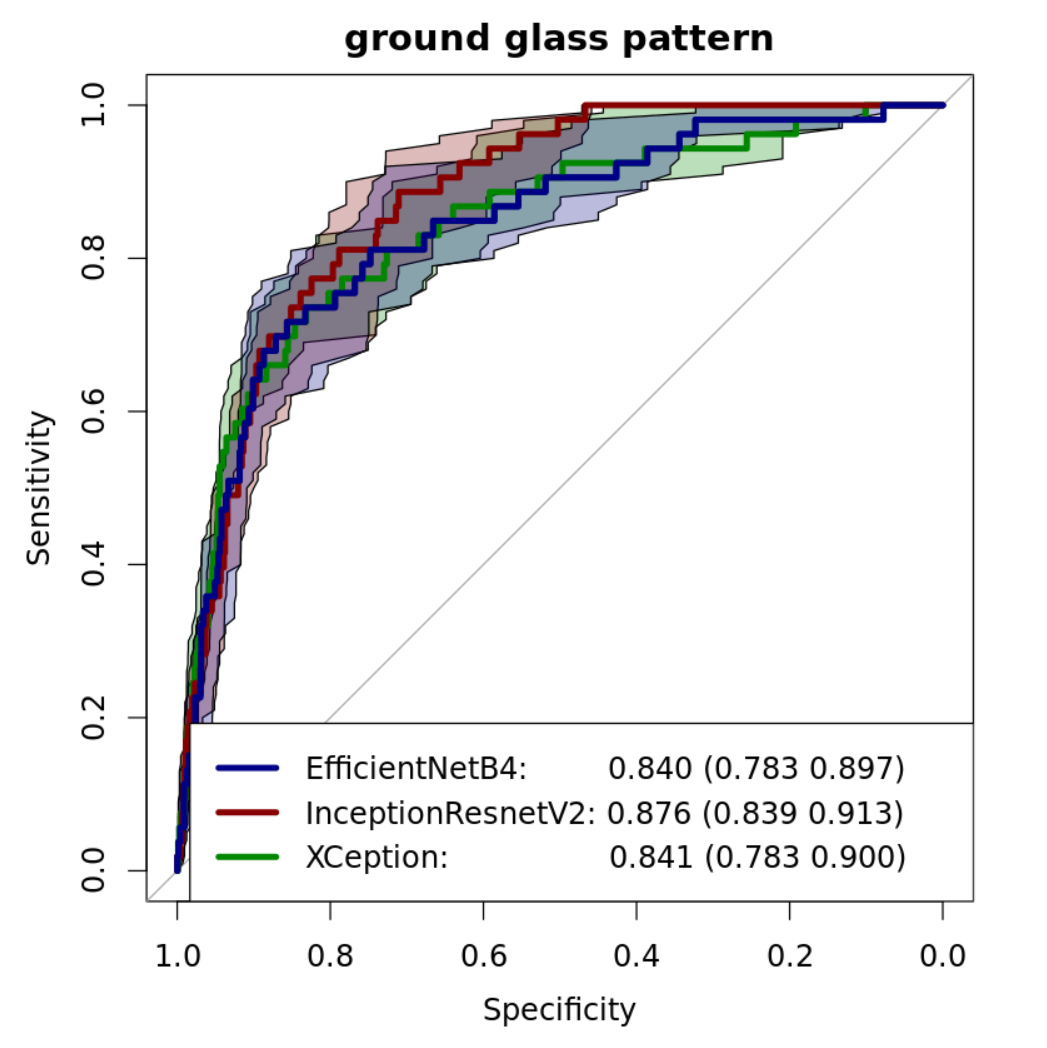} & 
        \includegraphics[trim={0 0 0cm 0}, width=0.45\textwidth]{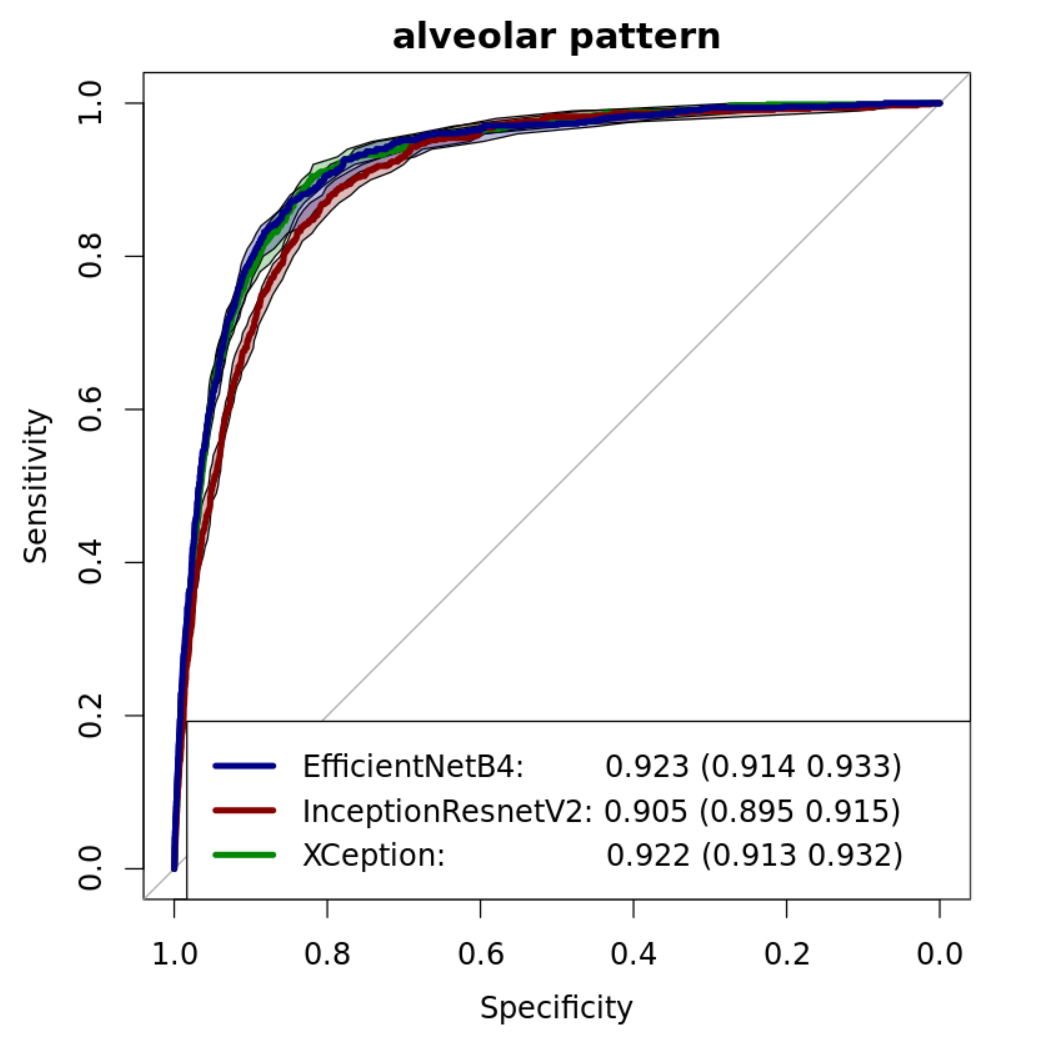} \\
  \end{tabular}
    
    \caption{Receiver operator characteristic curves for COVID-19 diagnosis and the radiological findings related to COVID-19 for the three network structures evaluated. Curves obtained on the test set. AUCs and 95\% confidence intervals are shown as shaded areas and quantified in the legend. }
    \label{fig:aucs}
\end{figure}

The aim of this work is to assist in the diagnosis of COVID-19. We therefore put special focus on the evaluation of those findings that are compatible with the disease. In chest x-rays, COVID-19 is associated with ground glass opacities, consolidations and alveolar patterns.  Fig~\ref{fig:aucs} shows the performance of the classifier for such variables. The area under the curve for the EfficientNetB4 network for COVID-19 is of 0.937 (95\% CI between 0.924 to 0.950), for consolidations is of 0.924 (95\% CI between 0.910 and 0.938), for ground glass opacities is of 0.840 (95\% CI interval between 0.783 and 0.897), for alveolar patterns is of 0.923 (95\% CI between 0.914 and 0.933) and for infiltrates of 0.871 (95\% CI between 0.863 and 0.879).

Of interest is to show the capacity of the model to distinguish between COVID-19 and other pneumonias. To perform such test we have selected the images from the test set that have any type of pneumonia and evaluate in such subsets the performance of COVID-19 detection. The AUCs obtained are 0.73 for EfficientnetB4 and 0.70 for both InceptionResnetV2 and XCeption. When evaluating the EfficientNetB4, no image from padchest had a score for COVID-19 higher than 0.5, showing that the network does not produce false positives on COVID-19 negative images.    

\begin{figure}[h]
    \centering
    \includegraphics[trim={6.5cm 2cm 5cm 0.5cm}, width=\columnwidth]{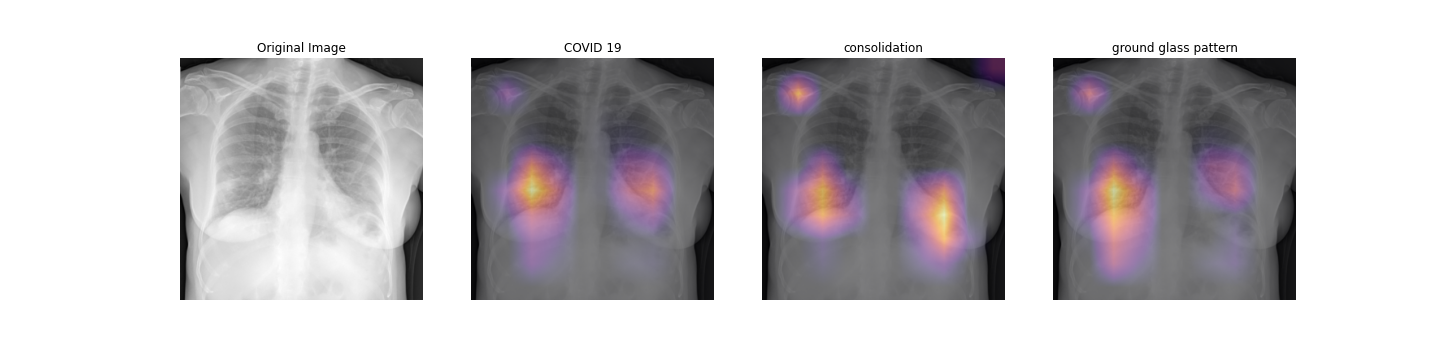}
    \caption{GradCam network attention heatmaps for a subject with COVID 19 pneumonia. It is the same subject as Fig.~\ref{fig:example}. Please note how the COVID 19 diagnosis is related to bilateral lower attention, the consolidation is associated with mostly left (anatomically) attention, while the ground glass pattern is associated mainly with  right attention.}
    \label{fig:gradcam}
\end{figure}

We have used the well known GradCam \cite{gradCAM}  method to estimate the regions of the image that contributed most to the prediction of the variables. An example is shown in Fig.~\ref{fig:gradcam}. Consolidations appear mainly at the left lung, ground glass opacities at the right lung and the diagnosis of COVID 19 is produced by features located at both the left and right lungs, mainly on the lower lobes. While the three heatmaps are obtained independently, thy complement each other and explain the decisions made by the network.

The software was implemented with Keras and Tensorflow libraries in Python. Training was performed in parallel on four Nvidia 1080Ti GPUs cards. It took 11.4 minutes per epoch for the EfficientNetB4 network, 11 minutes per epoch for InceptionResnetV2 network and 21 minutes for XCeption. 



\section{Discussion}
\label{sec:discussion}

In this work, we present a method for chest x-ray image 
classification based on deep neural networks which outputs hierarchical radiological findings, differential diagnosis and localization outcome variables. The reference standard is obtained using natural language processing from the radiological reports. 

On the 23,159 test images, the method shows a high average AUC score of all the variables (0.79 for EfficientNetB4), holding promise for its future use. Radiological findings that are large in size, specially due to artificial objects such as pacemakers or heart valves, are the ones showing higher AUCs. Smaller or subtle findings such as azygos lobe, rib fractures or air trapping show significantly lower performance. This could be partly explained by the image size that has been used in the method. Due to computational reasons, the image size had to be reduced to 299x299 pixels. Full-size images are around 8 megapixels. Such reduction in the image resolution eliminates textures and edges, that could be of great importance to detect subtle structures.

When evaluating the performance on COVID-19, the proposed method successfully distinguished COVID-19 patients, with an AUC of $0.94$, showing promising results for the diagnosis of cases amidst the pandemic. When evaluated to distinguish between COVID-19 caused pneumonia and other pneumonia types, the AUC dropped to $0.73$. When evaluating the network on the subset of patients with pneumonia from the PadChest dataset, no subject had a COVID-19 score higher than $0.5$. The drop in the AUC can be explained by several reasons. The first one is the source of the reference standard. The BIMCV COVID-19+ dataset has been acquired amidst the beginning of the COVID-19 pandemic and the radiological reports may not have correctly included COVID-19 information. For instance, 10 out of the 18 image studies with regions of interest included in the dataset were not marked as being COVID-19 positive in the radiology report, although they have ground glass opacities or consolidations that were marked by reviewing radiologists specialized in COVID-19. 

The reference standard used for training comes from natural language processing methods applied to the radiological reports. As mentioned by \cite{Bluemke2020}, such reference standard is considered of \textit{moderate quality}, and task-specific expert annotation is preferred for the training and development of artificial intelligence methods. The source of the reference standard is a limitation of the proposed work. It should be note that all prior art related to automated interpretation of chest x-rays using the datasets of~\cite{irvin2019chexpert,bustos2019padchest,johnson2019mimic,wang2017chestx} have the same limitation. An evaluation of the quality of the reference standard would be necessary to solidify the results. 

Further, the reference standard is composed of disease and locations, which are classified independently. It would be of interest to intertwine them, such that diseases that are in certain locations can not produce outcomes in incompatible locations and vice-versa. For instance, cardiomegaly as a single radiological finding can not appear at the same time as the location peripheral. Beyond locations, the hierarchical relationships of the variables are exploited at training, but not at detection, and they could be used to make stronger predictions. For instance, a moderately positive ground glass pattern could be rejected if their parent neurons did not fire adequately. Such methods are left as future work.

Different image classification networks can be used for the proposed methodology. In this work we have evaluated XCeption, InceptionResnetV2 and EFficientNetB4. The performance order followed that of the original networks on the ImageNet classification database, being EfficientNetB4 the network that has a  highest average AUC. Still, the network and the training methodology chosen in this work are relatively simple for current state-of-the-art deep neural networks. Methods such as 
ensemble learning and cosine decay learning rate~\cite{Karim2020}, uncertainty estimation ~\cite{Ghoshal2020} or more advanced hierarchical training methods may improve the results. 

A topic of discussion is whether the construction of the prediction labels by placing a '1' on the labels in the reference standard as well an all its parents is correct. For instance, while ground glass opacity (GGO) is an infiltrate, when describing the radiological findings in one image, one should clearly state that it is a GGO and not an infiltrate. We have addressed such topic in a two-fold manner. For training, we place '1' on all variables until the root. We do so since, using the same example as before, all GGO are indeed infiltrates. However, when we perform testing for GGO, all images that does not have GGO in the reference standard, even those with other kind of infiltrates, are marked as '0'. This allows us to accurately evaluate the performance of GGO detection.

While the two datasets mixed for training are from the same region of Spain, they differ in their origin. Padchest is a single institution dataset (images were obtained from one hospital), while BIMCV COVID-19+ images were obtained from several institutions. This may lead to domain subtleties that are guiding the performance. A more in-depth evaluation of whether such is the case would be needed and is a limitation of this work. Furthermore, it should be considered that the majority of the population in the Valencian region is Caucasian. 

Despite the above-mentioned limitations, the proposed work has value, since, to our knowledge, it is the first neural network used to interpret chest x-rays that uses hierarchical labels based on the UMLS entity relationships of radiological findings, differential diagnoses and locations. Further, it is the first work to use a large, open, dataset of COVID-19+ images for training and evaluation that comes from the same sources as the COVID-19 negative samples.

\section{Conclusions}
\label{sec:conclusions}

We have presented a method for the detection of hierarchical radiological findings (189 nodes), their anatomical location (122 nodes) and the diagnoses associated (22 variables) from chest x-rays. The method has been trained and evaluated on a composed large dataset of 94,659 frontal chest x-rays, achieving an average AUC of 0.787 for all the variables. When evaluated on COVID-19+ cases, results show a reliable performance, with an AUC of 0.94. The main advantage of the proposed method compared to previous works is that it provides explainability of the diagnoses, since not only the outcome variable are given, but also the radiological findings that have led to diagnoses, such as ground glass patterns (AUC of 0.84), consolidations (AUC of 0.92) or alveolar patterns (AUC of 0.92). 

We believe that the proposed hierarchical labeling and training could be applied for further work on chest x-ray image interpretation. It is also planned to annotate COVID-19 related findings at a pixel level in order to perform semantic segmentation, since the number and area of these lesions is relevant both for the diagnosis and the prognostic.

\section*{Acknowledgments}
This work has been supported by the grant awarded through decree 51/2020 by the Valencian Innovation Agency (Spain). This research is also supported by the project UACOVID19-18 from the University of Alicante. We thank also NVIDIA for the generous donation of a Titan Xp and a Quadro P6000 used in this research.

\bibliographystyle{splncs04} 
\bibliography{refs}

\begin{thebibliography}{10}
\providecommand{\url}[1]{\texttt{#1}}
\providecommand{\urlprefix}{URL }
\providecommand{\doi}[1]{https://doi.org/#1}

\bibitem{Apostolopoulos2020}
Apostolopoulos, I.D., Mpesiana, T.A.: {Covid-19: automatic detection from X-ray
  images utilizing transfer learning with convolutional neural networks}.
  Physical and Engineering Sciences in Medicine (0123456789), ~1--6 (2020).
  \doi{10.1007/s13246-020-00865-4}

\bibitem{Bernheim20}
Bernheim, A., Mei, X., Huang, M., Yang, Y., Fayad, Z., Zhang, N., Diao, K.,
  Lin, B., Zhu, X., Li, K., Li, S., Shan, H., Jacobi, A., Chung, M.: Chest ct
  findings in coronavirus disease-19 (covid-19): Relationship to duration of
  infection. Radiology  \textbf{295},  200463 (02 2020).
  \doi{10.1148/radiol.2020200463}

\bibitem{Bluemke2020}
Bluemke, D.A., Moy, L., Bredella, M.A., Ertl-Wagner, B.B., Fowler, K.J., Goh,
  V.J., Halpern, E.F., Hess, C.P., Schiebler, M.L., Weiss, C.R.: {Assessing
  Radiology Research on Artificial Intelligence: A Brief Guide for Authors,
  Reviewers, and Readers-From the Radiology Editorial Board Key Considerations
  for Authors, Reviewers, and Readers of AI/ML Manuscripts in Radiology}.
  Radiology  \textbf{294} (2020). \doi{10.1148/radiol.2019192515},
  \url{https://doi.org/10.1148/radiol.2019192515}

\bibitem{xnet}
Bullock, J., Cuesta-L{\'a}zaro, C., Quera-Bofarull, A.: Xnet: A convolutional
  neural network (cnn) implementation for medical x-ray image segmentation
  suitable for small datasets. In: Medical Imaging 2019: Biomedical
  Applications in Molecular, Structural, and Functional Imaging. vol. 10953, p.
  109531Z. International Society for Optics and Photonics (2019)

\bibitem{Burlacu2020}
Burlacu, A., Crisan-Dabija, R., Popa, I.V., Artene, B., Birzu, V., Pricop, M.,
  Plesoianu, C., Generali, D.: Curbing the ai-induced enthusiasm in diagnosing
  covid-19 on chest x-rays: the present and the near-future. medRxiv  (2020)

\bibitem{bustos2019padchest}
Bustos, A., Pertusa, A., Salinas, J.M., de~la Iglesia-Vay{\'a}, M.: Padchest: A
  large chest x-ray image dataset with multi-label annotated reports. arXiv
  preprint arXiv:1901.07441  (2019)

\bibitem{Chowdhury2020}
Chowdhury, M.E., Rahman, T., Khandakar, A., Mazhar, R., Kadir, M.A., Mahbub,
  Z.B., Islam, K.R., Khan, M.S., Iqbal, A., Al-Emadi, N., et~al.: Can ai help
  in screening viral and covid-19 pneumonia? arXiv preprint arXiv:2003.13145
  (2020)

\bibitem{Cohen2020}
Cohen, J.P., Morrison, P., Dao, L.: {COVID-19 Image Data Collection} (2020)

\bibitem{imagenet}
Deng, J., Dong, W., Socher, R., Li, L.J., Li, K., Fei-Fei, L.: Imagenet: A
  large-scale hierarchical image database. In: 2009 IEEE conference on computer
  vision and pattern recognition. pp. 248--255. Ieee (2009)

\bibitem{Farooq2020}
Farooq, M., Hafeez, A.: {COVID-ResNet: A Deep Learning Framework for Screening
  of COVID19 from Radiographs}  (2020), \url{http://arxiv.org/abs/2003.14395}

\bibitem{Ghoshal2020}
Ghoshal, B., Tucker, A.: {Estimating Uncertainty and Interpretability in Deep
  Learning for Coronavirus (COVID-19) Detection} pp. 1--14 (2020),
  \url{http://arxiv.org/abs/2003.10769}

\bibitem{BIMCVCOVID19}
de~la Iglesia~Vayá, M., Saborit, J.M., Montell, J.A., Pertusa, A., Bustos, A.,
  Cazorla, M., Galant, J., Barber, X., Orozco-Beltrán, D., Garcia, F.,
  Caparrós, M., González, G., Salinas, J.M.: Bimcv covid-19+: a large
  annotated dataset of rx and ct images from covid-19 patients (2020),
  \url{http://bimcv.cipf.es/bimcv-projects/bimcv-covid19/}

\bibitem{irvin2019chexpert}
Irvin, J., Rajpurkar, P., Ko, M., Yu, Y., Ciurea-Ilcus, S., Chute, C.,
  Marklund, H., Haghgoo, B., Ball, R., Shpanskaya, K., et~al.: Chexpert: A
  large chest radiograph dataset with uncertainty labels and expert comparison.
  In: Proceedings of the AAAI Conference on Artificial Intelligence. vol.~33,
  pp. 590--597 (2019)

\bibitem{johnson2019mimic}
Johnson, A.E., Pollard, T.J., Berkowitz, S., Greenbaum, N.R., Lungren, M.P.,
  Deng, C.y., Mark, R.G., Horng, S.: Mimic-cxr: A large publicly available
  database of labeled chest radiographs. arXiv preprint arXiv:1901.07042
  (2019)

\bibitem{Karim2020}
Karim, M., D{\"o}hmen, T., Rebholz-Schuhmann, D., Decker, S., Cochez, M.,
  Beyan, O., et~al.: Deepcovidexplainer: Explainable covid-19 predictions based
  on chest x-ray images. arXiv preprint arXiv:2004.04582  (2020)

\bibitem{Kim2020}
Kim, H., Hong, H., Yoon, S.H.: Diagnostic performance of ct and reverse
  transcriptase-polymerase chain reaction for coronavirus disease 2019: a
  meta-analysis. Radiology p. 201343 (2020)

\bibitem{Latif2020}
Latif, S., Usman, M., Manzoor, S., Iqbal, W., Qadir, J., Tyson, G., Castro, I.,
  Razi, A., Boulos, M.N.K., Weller, A., et~al.: Leveraging data science to
  combat covid-19: A comprehensive review  (2020)

\bibitem{Li2020}
Li, X., Li, C., Zhu, D.: {COVID-MobileXpert: On-Device COVID-19 Screening using
  Snapshots of Chest X-Ray}  (2020), \url{http://arxiv.org/abs/2004.03042}

\bibitem{Mei-sze2020}
Ng, M.Y., Lee, E.Y., Yang, J., Yang, F., Li, X., Wang, H., Lui, M.M.s., Lo,
  C.S.Y., Leung, B., Khong, P.L., et~al.: Imaging profile of the covid-19
  infection: radiologic findings and literature review. Radiology:
  Cardiothoracic Imaging  \textbf{2}(1),  e200034 (2020)

\bibitem{Pan20}
Pan, F., Ye, T., Sun, P., Gui, S., Liang, B., Li, L., Zheng, D., Wang, J.,
  Hesketh, R., Yang, L., Zheng, C.: Time course of lung changes on chest ct
  during recovery from 2019 novel coronavirus (covid-19) pneumonia. Radiology
  \textbf{295},  200370 (02 2020). \doi{10.1148/radiol.2020200370}

\bibitem{Kaggle}
{Paul Mooney}: {Chest X-Ray Images (Pneumonia)},
  \url{https://www.kaggle.com/paultimothymooney/chest-xray-pneumonia}

\bibitem{RSNA}
{Radiological Society of North America}: {RSNA Challenge},
  \url{https://www.rsna.org/en/education/ai-resources-and-training/ai-image-challenge}

\bibitem{rajpurkar2017chexnet}
Rajpurkar, P., Irvin, J., Zhu, K., Yang, B., Mehta, H., Duan, T., Ding, D.,
  Bagul, A., Langlotz, C., Shpanskaya, K., Lungren, M.P., Ng, A.Y.: Chexnet:
  Radiologist-level pneumonia detection on chest x-rays with deep learning.
  arXiv preprint arXiv:1711.05225  (2017)

\bibitem{gradCAM}
Selvaraju, R.R., Cogswell, M., Das, A., Vedantam, R., Parikh, D., Batra, D.:
  Grad-cam: Visual explanations from deep networks via gradient-based
  localization. In: The IEEE International Conference on Computer Vision (ICCV)
  (Oct 2017)

\bibitem{ShiFeng2020}
Shi, F., Wang, J., Shi, J., Wu, Z., Wang, Q., Tang, Z., He, K., Shi, Y., Shen,
  D.: Review of artificial intelligence techniques in imaging data acquisition,
  segmentation and diagnosis for covid-19. IEEE Reviews in Biomedical
  Engineering  (2020)

\bibitem{Shi2020}
Shi, H., Han, X., Jiang, N., Cao, Y., Alwalid, O., Gu, J., Fan, Y., Zheng, C.:
  Radiological findings from 81 patients with covid-19 pneumonia in wuhan,
  china: a descriptive study. The Lancet Infectious Diseases  (2020)

\bibitem{inceptionresnetv2}
Szegedy, C., Ioffe, S., Vanhoucke, V., Alemi, A.A.: {Inception-v4,
  Inception-ResNet and the Impact of Residual Connections on Learning}. In:
  {Proceedings of the Thirty-First AAAI Conference on Artificial Intelligence}.
  p. 4278–4284. AAAI’17, AAAI Press (2017)

\bibitem{effnet}
Tan, M., Le, Q.V.: Efficientnet: Rethinking model scaling for convolutional
  neural networks. In: Chaudhuri, K., Salakhutdinov, R. (eds.) Proceedings of
  the 36th International Conference on Machine Learning, {ICML} 2019, 9-15 June
  2019, Long Beach, California, {USA}. Proceedings of Machine Learning
  Research, vol.~97, pp. 6105--6114. {PMLR} (2019),
  \url{http://proceedings.mlr.press/v97/tan19a.html}

\bibitem{varshni19}
{Varshni}, D., {Thakral}, K., {Agarwal}, L., {Nijhawan}, R., {Mittal}, A.:
  Pneumonia detection using cnn based feature extraction. In: 2019 IEEE
  International Conference on Electrical, Computer and Communication
  Technologies (ICECCT). pp.~1--7 (2019)

\bibitem{Wang2020}
Wang, L., Wong, A.: {COVID-Net: A Tailored Deep Convolutional Neural Network
  Design for Detection of COVID-19 Cases from Chest X-Ray Images} pp. 1--12
  (2020), \url{http://arxiv.org/abs/2003.09871}

\bibitem{wang2017chestx}
Wang, X., Peng, Y., Lu, L., Lu, Z., Bagheri, M., Summers, R.M.: Chestx-ray8:
  Hospital-scale chest x-ray database and benchmarks on weakly-supervised
  classification and localization of common thorax diseases. In: Computer
  Vision and Pattern Recognition (CVPR), 2017 IEEE Conference on. pp.
  3462--3471. IEEE (2017)

\bibitem{Wang20}
Wang, Y., Dong, C., Hu, Y., Li, C., Ren, Q., Zhang, X., Shi, H., Zhou, M.:
  Temporal changes of ct findings in 90 patients with covid-19 pneumonia: A
  longitudinal study. Radiology p. 200843 (03 2020).
  \doi{10.1148/radiol.2020200843}

\bibitem{YuenFrankWong2020}
Wong, H.Y.F., Lam, H.Y.S., Fong, A.H.T., Leung, S.T., Chin, T.W.Y., Lo, C.S.Y.,
  Lui, M.M.S., Lee, J.C.Y., Chiu, K.W.H., Chung, T., et~al.: Frequency and
  distribution of chest radiographic findings in covid-19 positive patients.
  Radiology p. 201160 (2020)

\end{thebibliography}

\section*{Supplementary Material}
\tiny
\begin{longtable}{|l|c|c|c|}
\hline
    \centering
  
Label  & EfficientNetBa4 & InceptionResnetV2 & XCeption\\
\hline
normal  [C0205307] [8412] & 0.86 & 0.83 & 0.86\\
exclude   [197] & 0.55 & 0.46 & 0.49\\
suboptimal study  [C2828075] [201] & 0.80 & 0.78 & 0.77\\
radiological finding  [C0436485] [12877] & 0.85 & 0.81 & 0.84\\
├── unchanged   [1713] & 0.70 & 0.67 & 0.68\\
├── obesity  [C0028754] [5] & 0.84 & 0.68 & 0.65\\
├── chronic changes  [C0742362] [978] & 0.79 & 0.76 & 0.78\\
├── calcified densities  [C2203586] [199] & 0.71 & 0.68 & 0.70\\
│\verb!   !├── calcified granuloma  [C0333404] [458] & 0.82 & 0.73 & 0.80\\
│\verb!   !├── calcified adenopathy   [83] & 0.72 & 0.69 & 0.73\\
│\verb!   !│\verb!   !└── calcified mediastinal adenopathy   [3] & 0.56 & 0.68 & 0.68\\
│\verb!   !├── calcified pleural thickening   [88] & 0.93 & 0.89 & 0.93\\
│\verb!   !├── calcified pleural plaques   [34] & 0.91 & 0.87 & 0.91\\
│\verb!   !├── heart valve calcified  [C2073448] [23] & 0.88 & 0.88 & 0.87\\
│\verb!   !└── calcified fibroadenoma   [10] & 0.80 & 0.63 & 0.66\\
├── granuloma  [C0235557] [557] & 0.82 & 0.72 & 0.79\\
│\verb!   !└── calcified granuloma  [C0333404] [458] & 0.82 & 0.73 & 0.80\\
├── end on vessel   [37] & 0.58 & 0.64 & 0.68\\
├── adenopathy  [C0478664] [131] & 0.69 & 0.66 & 0.68\\
│\verb!   !└── calcified adenopathy   [83] & 0.72 & 0.69 & 0.73\\
├── nodule  [C0034079] [576] & 0.75 & 0.70 & 0.73\\
│\verb!   !└── multiple nodules  [C2073563] [45] & 0.92 & 0.86 & 0.88\\
├── pseudonodule   [438] & 0.73 & 0.68 & 0.73\\
│\verb!   !├── nipple shadow   [231] & 0.83 & 0.72 & 0.79\\
│\verb!   !└── end on vessel   [37] & 0.58 & 0.64 & 0.68\\
├── abscess  [C0024110] [1] & 0.98 & 0.78 & 0.91\\
├── cyst  [C2073485] [1] & 0.81 & 0.96 & 0.96\\
├── cavitation  [C0578537] [77] & 0.94 & 0.94 & 0.95\\
├── fibrotic band  [C0865843] [538] & 0.86 & 0.80 & 0.84\\
├── volume loss  [C3203358] [363] & 0.92 & 0.89 & 0.92\\
├── hypoexpansion   [144] & 0.83 & 0.78 & 0.76\\
│\verb!   !└── hypoexpansion basal   [19] & 0.84 & 0.72 & 0.74\\
├── bullas  [C0241982] [92] & 0.87 & 0.86 & 0.85\\
├── pneumothorax  [C2073565] [59] & 0.77 & 0.71 & 0.75\\
│\verb!   !└── hydropneumothorax  [C0020303] [5] & 0.93 & 0.85 & 0.90\\
├── pneumoperitoneo  [C0032320] [7] & 0.79 & 0.64 & 0.80\\
├── pneumomediastinum  [C2073636] [2] & 0.44 & 0.25 & 0.58\\
├── subcutaneous emphysema  [C0038536] [8] & 0.93 & 0.91 & 0.94\\
├── hyperinflated lung  [C0546312] [94] & 0.80 & 0.75 & 0.73\\
├── flattened diaphragm  [C2073504] [136] & 0.74 & 0.70 & 0.72\\
├── lung vascular paucity   [31] & 0.73 & 0.72 & 0.72\\
├── air trapping  [C0231819] [798] & 0.75 & 0.71 & 0.72\\
├── bronchiectasis  [C0006267] [363] & 0.81 & 0.78 & 0.80\\
├── infiltrates  [C0277877] [2103] & 0.87 & 0.84 & 0.86\\
│\verb!   !├── interstitial pattern  [C2073538] [907] & 0.85 & 0.81 & 0.82\\
│\verb!   !│\verb!   !├── ground glass pattern  [C3544344] [53] & 0.84 & 0.88 & 0.84\\
│\verb!   !│\verb!   !├── reticular interstitial pattern   [96] & 0.82 & 0.74 & 0.74\\
│\verb!   !│\verb!   !├── reticulonodular interstitial pattern  [C2073672] [38] & 0.86 & 0.83 & 0.82\\
│\verb!   !│\verb!   !└── miliary opacities  [C2073583] [25] & 0.72 & 0.73 & 0.72\\
│\verb!   !└── alveolar pattern  [C1332240] [683] & 0.92 & 0.91 & 0.92\\
│\verb!   !\verb!   ! ├── consolidation  [C0521530] [315] & 0.92 & 0.90 & 0.92\\
│\verb!   !\verb!   ! │\verb!   !└── air bronchogram  [C3669021] [16] & 0.77 & 0.76 & 0.70\\
│\verb!   !\verb!   ! └── air bronchogram  [C3669021] [16] & 0.77 & 0.76 & 0.70\\
├── bronchovascular markings  [C2073518] [182] & 0.82 & 0.75 & 0.80\\
├── air fluid level  [C0740844] [24] & 0.75 & 0.75 & 0.71\\
├── increased density  [C1443940] [540] & 0.80 & 0.78 & 0.78\\
├── atelectasis  [C0004144] [1002] & 0.85 & 0.74 & 0.81\\
│\verb!   !├── total atelectasis  [C0264497] [2] & 0.99 & 0.94 & 0.85\\
│\verb!   !├── lobar atelectasis   [88] & 0.89 & 0.83 & 0.87\\
│\verb!   !├── segmental atelectasis   [21] & 0.73 & 0.77 & 0.72\\
│\verb!   !├── laminar atelectasis   [698] & 0.84 & 0.70 & 0.78\\
│\verb!   !├── round atelectasis  [C2062952] [1] & 0.93 & 0.91 & 0.94\\
│\verb!   !└── atelectasis basal  [C0746053] [3] & 0.95 & 0.78 & 0.86\\
├── mediastinal shift  [C0264576] [9] & 0.73 & 0.75 & 0.69\\
├── azygos lobe  [C0265794] [74] & 0.56 & 0.61 & 0.62\\
├── fissure thickening   [80] & 0.89 & 0.78 & 0.84\\
│\verb!   !├── minor fissure thickening   [61] & 0.90 & 0.77 & 0.84\\
│\verb!   !├── major fissure thickening   [10] & 0.78 & 0.68 & 0.76\\
│\verb!   !└── loculated fissural effusion   [20] & 0.91 & 0.86 & 0.89\\
├── pleural thickening  [C0264545] [789] & 0.85 & 0.80 & 0.83\\
│\verb!   !├── apical pleural thickening   [585] & 0.86 & 0.80 & 0.84\\
│\verb!   !└── calcified pleural thickening   [88] & 0.93 & 0.89 & 0.93\\
├── pleural plaques  [C0340030] [38] & 0.88 & 0.85 & 0.88\\
│\verb!   !└── calcified pleural plaques   [34] & 0.91 & 0.87 & 0.91\\
├── pleural effusion  [C2073625] [786] & 0.96 & 0.91 & 0.95\\
│\verb!   !├── loculated pleural effusion  [C0747639] [29] & 0.95 & 0.92 & 0.96\\
│\verb!   !├── loculated fissural effusion   [20] & 0.91 & 0.86 & 0.89\\
│\verb!   !├── hydropneumothorax  [C0020303] [5] & 0.93 & 0.85 & 0.90\\
│\verb!   !├── empyema  [C0014009] [0] & 0.00 & 0.00 & 0.00\\
│\verb!   !└── hemothorax  [C0019123] [0] & 0.00 & 0.00 & 0.00\\
├── pleural mass  [C1709576] [0] & 0.00 & 0.00 & 0.00\\
├── costophrenic angle blunting  [C0742855] [810] & 0.89 & 0.78 & 0.85\\
├── vascular redistribution  [C0239041] [64] & 0.88 & 0.85 & 0.87\\
│\verb!   !└── central vascular redistribution   [14] & 0.83 & 0.83 & 0.88\\
├── hilar enlargement  [C1698506] [868] & 0.73 & 0.68 & 0.71\\
│\verb!   !├── adenopathy  [C0149711] [131] & 0.69 & 0.66 & 0.68\\
│\verb!   !└── vascular hilar enlargement   [602] & 0.76 & 0.68 & 0.73\\
│\verb!   !\verb!   ! └── pulmonary artery enlargement  [C2072932] [36] & 0.69 & 0.64 & 0.63\\
├── hilar congestion  [C0582411] [80] & 0.88 & 0.89 & 0.88\\
├── cardiomegaly  [C0018800] [2068] & 0.93 & 0.91 & 0.92\\
├── pericardial effusion  [C0031039] [11] & 0.90 & 0.92 & 0.79\\
├── kerley lines  [C0239019] [18] & 0.81 & 0.79 & 0.80\\
├── dextrocardia  [C0011813] [1] & 0.86 & 0.98 & 0.96\\
├── right sided aortic arch  [C0035615] [7] & 0.74 & 0.57 & 0.71\\
├── aortic atheromatosis  [C1096249] [364] & 0.83 & 0.81 & 0.81\\
├── aortic elongation   [1728] & 0.90 & 0.87 & 0.89\\
│\verb!   !├── descendent aortic elongation  [C4476542] [174] & 0.87 & 0.82 & 0.84\\
│\verb!   !├── ascendent aortic elongation  [C3889085] [31] & 0.79 & 0.78 & 0.74\\
│\verb!   !├── aortic button enlargement  [C1851119] [76] & 0.78 & 0.75 & 0.78\\
│\verb!   !└── supra aortic elongation   [238] & 0.87 & 0.84 & 0.86\\
├── aortic aneurysm  [C0003486] [7] & 0.94 & 0.74 & 0.89\\
├── mediastinal enlargement  [C2021206] [646] & 0.82 & 0.79 & 0.81\\
│\verb!   !├── superior mediastinal enlargement  [C4273001] [414] & 0.82 & 0.78 & 0.81\\
│\verb!   !│\verb!   !├── goiter  [C0018021] [175] & 0.78 & 0.73 & 0.80\\
│\verb!   !│\verb!   !└── supra aortic elongation   [238] & 0.87 & 0.84 & 0.86\\
│\verb!   !├── descendent aortic elongation  [C4476542] [174] & 0.87 & 0.82 & 0.84\\
│\verb!   !├── ascendent aortic elongation  [C3889085] [31] & 0.79 & 0.78 & 0.74\\
│\verb!   !├── aortic aneurysm  [C0003486] [7] & 0.94 & 0.74 & 0.89\\
│\verb!   !├── mediastinal mass  [C0240318] [41] & 0.73 & 0.67 & 0.73\\
│\verb!   !└── hiatal hernia  [C3489393] [368] & 0.87 & 0.80 & 0.85\\
├── tracheal shift   [148] & 0.76 & 0.66 & 0.79\\
├── mass  [C2603353] [175] & 0.80 & 0.76 & 0.79\\
│\verb!   !├── mediastinal mass  [C0240318] [41] & 0.73 & 0.67 & 0.73\\
│\verb!   !├── breast mass  [C0024103] [0] & 0.00 & 0.00 & 0.00\\
│\verb!   !├── pleural mass  [C1709576] [0] & 0.00 & 0.00 & 0.00\\
│\verb!   !├── pulmonary mass  [C0149726] [117] & 0.90 & 0.86 & 0.90\\
│\verb!   !└── soft tissue mass  [C0457196] [16] & 0.57 & 0.63 & 0.51\\
├── esophagic dilatation  [C0192389] [1] & 0.90 & 0.85 & 0.86\\
├── azygoesophageal recess shift   [7] & 0.65 & 0.63 & 0.55\\
├── pericardial effusion  [C0031039] [11] & 0.90 & 0.92 & 0.79\\
├── lipomatosis   [76] & 0.71 & 0.70 & 0.64\\
│\verb!   !└── mediastinic lipomatosis  [C1333298] [76] & 0.71 & 0.69 & 0.65\\
├── thoracic cage deformation  [C4538889] [1847] & 0.80 & 0.67 & 0.73\\
│\verb!   !├── scoliosis  [C0036439] [1264] & 0.82 & 0.67 & 0.74\\
│\verb!   !├── kyphosis  [C2115817] [602] & 0.81 & 0.77 & 0.78\\
│\verb!   !├── pectum excavatum  [C2051831] [47] & 0.77 & 0.72 & 0.69\\
│\verb!   !├── pectum carinatum  [C2939416] [3] & 0.58 & 0.47 & 0.63\\
│\verb!   !└── cervical rib  [C0158779] [16] & 0.75 & 0.62 & 0.76\\
├── vertebral degenerative changes  [C4290224] [934] & 0.74 & 0.72 & 0.72\\
│\verb!   !└── vertebral compression  [C0262431] [290] & 0.76 & 0.72 & 0.74\\
│\verb!   !\verb!   ! └── vertebral anterior compression   [253] & 0.75 & 0.72 & 0.75\\
├── lytic bone lesion  [C0476382] [12] & 0.73 & 0.75 & 0.71\\
├── sclerotic bone lesion  [C4315325] [93] & 0.65 & 0.58 & 0.62\\
│\verb!   !└── blastic bone lesion  [C2203581] [9] & 0.83 & 0.74 & 0.73\\
├── costochondral junction hypertrophy   [17] & 0.62 & 0.57 & 0.64\\
├── sternoclavicular junction hypertrophy   [1] & 0.31 & 0.16 & 0.09\\
├── axial hyperostosis  [C1400000] [34] & 0.71 & 0.70 & 0.66\\
├── osteopenia  [C0029453] [81] & 0.86 & 0.80 & 0.80\\
├── osteoporosis  [C0029456] [47] & 0.88 & 0.83 & 0.88\\
├── non axial articular degenerative changes   [34] & 0.84 & 0.81 & 0.81\\
├── subacromial space narrowing   [11] & 0.67 & 0.76 & 0.68\\
├── fracture  [C0016658] [591] & 0.75 & 0.69 & 0.73\\
│\verb!   !├── clavicle fracture  [C0159658] [38] & 0.57 & 0.56 & 0.55\\
│\verb!   !├── humeral fracture  [C0020162] [29] & 0.79 & 0.74 & 0.73\\
│\verb!   !├── vertebral fracture  [C0080179] [36] & 0.86 & 0.84 & 0.86\\
│\verb!   !└── rib fracture  [C0035522] [496] & 0.75 & 0.69 & 0.73\\
│\verb!   !\verb!   ! └── callus rib fracture  [C0006767] [414] & 0.76 & 0.71 & 0.74\\
├── gynecomastia  [C0018418] [107] & 0.82 & 0.72 & 0.79\\
├── hiatal hernia  [C3489393] [368] & 0.87 & 0.80 & 0.85\\
├── Chilaiditi sign  [C3178780] [1] & 0.93 & 0.91 & 0.89\\
├── hemidiaphragm elevation  [C2073707] [300] & 0.83 & 0.72 & 0.77\\
├── diaphragmatic eventration  [C0011981] [182] & 0.78 & 0.72 & 0.72\\
├── tracheostomy tube  [C0184159] [40] & 0.98 & 0.85 & 0.96\\
├── endotracheal tube  [C0336630] [16] & 0.99 & 0.92 & 0.95\\
├── NSG tube   [49] & 0.91 & 0.82 & 0.86\\
├── chest drain tube  [C0008034] [39] & 0.91 & 0.87 & 0.90\\
├── ventriculoperitoneal drain tube  [C0162702] [5] & 0.49 & 0.55 & 0.46\\
├── gastrostomy tube  [C0150595] [2] & 0.91 & 0.59 & 0.96\\
├── nephrostomy tube  [C0184149] [0] & 0.00 & 0.00 & 0.00\\
├── double J stent  [C0441293] [2] & 0.51 & 0.51 & 0.68\\
├── catheter  [C0085590] [137] & 0.92 & 0.86 & 0.92\\
│\verb!   !└── central venous catheter  [C1145640] [125] & 0.94 & 0.89 & 0.93\\
│\verb!   !\verb!   ! ├── cvc via subclavian vein  [C0398281] [22] & 0.92 & 0.85 & 0.91\\
│\verb!   !\verb!   ! ├── cvc via jugular vein  [C0398278] [18] & 0.91 & 0.93 & 0.88\\
│\verb!   !\verb!   ! ├── reservoir cvc  [C2026143] [66] & 0.96 & 0.91 & 0.97\\
│\verb!   !\verb!   ! └── cvc via umbilical vein  [C0398284] [0] & 0.00 & 0.00 & 0.00\\
├── electrical device   [406] & 0.99 & 0.99 & 0.99\\
│\verb!   !├── dual chamber device  [C2732817] [164] & 0.99 & 0.99 & 0.99\\
│\verb!   !├── single chamber device  [C2733312] [69] & 0.99 & 0.99 & 0.99\\
│\verb!   !├── pacemaker  [C0030163] [367] & 1.00 & 1.00 & 1.00\\
│\verb!   !└── dai  [C0972395] [36] & 0.98 & 0.96 & 0.96\\
├── artificial heart valve  [C1399223] [146] & 0.99 & 0.98 & 0.99\\
│\verb!   !├── artificial mitral heart valve  [C0869752] [56] & 0.99 & 0.98 & 0.99\\
│\verb!   !└── artificial aortic heart valve  [C0869748] [46] & 0.99 & 0.97 & 0.99\\
├── surgery   [995] & 0.89 & 0.78 & 0.85\\
│\verb!   !├── metal  [C0025552] [690] & 0.92 & 0.85 & 0.91\\
│\verb!   !│\verb!   !├── osteosynthesis material  [C0016642] [101] & 0.86 & 0.62 & 0.82\\
│\verb!   !│\verb!   !├── sternotomy  [C0185792] [382] & 1.00 & 0.97 & 1.00\\
│\verb!   !│\verb!   !└── suture material  [C4305366] [267] & 0.90 & 0.85 & 0.88\\
│\verb!   !├── bone cement  [C0005934] [0] & 0.00 & 0.00 & 0.00\\
│\verb!   !├── prosthesis  [C0175649] [117] & 0.84 & 0.66 & 0.77\\
│\verb!   !│\verb!   !├── humeral prosthesis  [C0745058] [12] & 0.97 & 0.75 & 0.94\\
│\verb!   !│\verb!   !├── mammary prosthesis  [C0917968] [95] & 0.85 & 0.70 & 0.77\\
│\verb!   !│\verb!   !└── endoprosthesis  [C0005846] [7] & 0.79 & 0.77 & 0.76\\
│\verb!   !│\verb!   !\verb!   ! └── aortic endoprosthesis   [2] & 0.95 & 0.92 & 0.99\\
│\verb!   !├── surgery breast  [C3714726] [175] & 0.83 & 0.70 & 0.75\\
│\verb!   !│\verb!   !└── mastectomy  [C0024881] [109] & 0.81 & 0.66 & 0.75\\
│\verb!   !├── surgery neck  [C0185773] [9] & 0.70 & 0.67 & 0.77\\
│\verb!   !├── surgery lung  [C0038903] [44] & 0.82 & 0.83 & 0.81\\
│\verb!   !├── surgery heart  [C0018821] [2] & 0.99 & 0.94 & 0.91\\
│\verb!   !└── surgery humeral  [C0186326] [2] & 0.99 & 0.86 & 0.93\\
├── abnormal foreign body  [C0016542] [4] & 0.94 & 0.84 & 0.94\\
└── external foreign body   [11] & 0.51 & 0.65 & 0.59\\
differential diagnosis   [4852] & 0.82 & 0.79 & 0.81\\
├── pneumonia  [C0032285] [1189] & 0.88 & 0.86 & 0.87\\
│\verb!   !└── atypical pneumonia  [C1412002] [340] & 0.93 & 0.92 & 0.91\\
│\verb!   !\verb!   ! └── viral pneumonia  [C0032310] [318] & 0.94 & 0.93 & 0.92\\
│\verb!   !\verb!   !\verb!   !  ├── COVID 19  [C5203670] [307] & 0.94 & 0.93 & 0.92\\
│\verb!   !\verb!   !\verb!   !  └── COVID 19 uncertain  [C5203671] [6] & 0.84 & 0.85 & 0.82\\
├── tuberculosis  [C0041296] [189] & 0.93 & 0.88 & 0.92\\
│\verb!   !└── tuberculosis sequelae  [C0494132] [148] & 0.95 & 0.91 & 0.94\\
├── lung metastasis  [C0153676] [49] & 0.91 & 0.87 & 0.86\\
├── lymphangitis carcinomatosa  [C0238258] [5] & 0.97 & 0.96 & 0.98\\
├── lepidic adenocarcinoma  [C4049711] [1] & 0.91 & 0.97 & 0.87\\
├── pulmonary fibrosis  [C0034069] [226] & 0.92 & 0.86 & 0.89\\
│\verb!   !├── post radiotherapy changes  [C1320687] [44] & 0.94 & 0.75 & 0.94\\
│\verb!   !└── asbestosis signs  [C0003949] [11] & 0.96 & 0.87 & 0.94\\
├── emphysema  [C0034067] [208] & 0.87 & 0.87 & 0.85\\
├── COPD signs  [C0024117] [3129] & 0.83 & 0.80 & 0.82\\
├── heart insufficiency  [C0018801] [163] & 0.94 & 0.93 & 0.92\\
├── respiratory distress  [C0476273] [0] & 0.00 & 0.00 & 0.00\\
├── pulmonary hypertension  [C0020542] [24] & 0.90 & 0.87 & 0.91\\
│\verb!   !├── pulmonary artery hypertension  [C2973725] [2] & 0.94 & 0.74 & 0.92\\
│\verb!   !└── pulmonary venous hypertension  [C4477098] [1] & 0.91 & 0.87 & 0.93\\
├── pulmonary edema  [C0034063] [53] & 0.96 & 0.96 & 0.96\\
└── bone metastasis  [C0153690] [23] & 0.86 & 0.80 & 0.83\\
localization   [12138] & 0.83 & 0.80 & 0.82\\
├── extracorporal  [C0424529] [0] & 0.00 & 0.00 & 0.00\\
├── cervical  [C0920882] [87] & 0.66 & 0.62 & 0.69\\
├── soft tissue  [C0225317] [726] & 0.77 & 0.66 & 0.71\\
│\verb!   !├── subcutaneous  [C0443315] [23] & 0.94 & 0.85 & 0.92\\
│\verb!   !├── axilar  [C0004454] [172] & 0.80 & 0.68 & 0.73\\
│\verb!   !└── pectoral  [C0230111] [468] & 0.81 & 0.67 & 0.76\\
│\verb!   !\verb!   ! └── nipple  [C0028109] [224] & 0.82 & 0.70 & 0.79\\
├── bone  [C0262950] [1727] & 0.70 & 0.66 & 0.67\\
│\verb!   !├── shoulder  [C0037004] [242] & 0.80 & 0.68 & 0.76\\
│\verb!   !│\verb!   !├── acromioclavicular joint  [C0001208] [16] & 0.70 & 0.57 & 0.62\\
│\verb!   !│\verb!   !├── rotator cuff  [C0085515] [17] & 0.59 & 0.58 & 0.61\\
│\verb!   !│\verb!   !├── supraspisnous  [C0225001] [9] & 0.70 & 0.73 & 0.62\\
│\verb!   !│\verb!   !└── humerus  [C0020164] [172] & 0.80 & 0.68 & 0.77\\
│\verb!   !│\verb!   !\verb!   ! ├── humeral head  [C0223683] [62] & 0.73 & 0.63 & 0.73\\
│\verb!   !│\verb!   !\verb!   ! ├── humeral neck  [C0448034] [13] & 0.79 & 0.71 & 0.69\\
│\verb!   !│\verb!   !\verb!   ! └── glenohumeral joint  [C0225063] [14] & 0.86 & 0.76 & 0.81\\
│\verb!   !├── clavicle  [C0008913] [65] & 0.59 & 0.52 & 0.57\\
│\verb!   !├── scapula  [C0036277] [43] & 0.65 & 0.53 & 0.64\\
│\verb!   !├── costoesternal  [C0450216] [9] & 0.60 & 0.63 & 0.61\\
│\verb!   !├── column  [C0037949] [599] & 0.71 & 0.70 & 0.69\\
│\verb!   !│\verb!   !├── intersomatic space  [C0223088] [15] & 0.64 & 0.63 & 0.67\\
│\verb!   !│\verb!   !├── dorsal vertebrae  [C0039987] [428] & 0.71 & 0.70 & 0.69\\
│\verb!   !│\verb!   !├── cervical vertebrae  [C3665420] [19] & 0.67 & 0.55 & 0.67\\
│\verb!   !│\verb!   !├── lumbar vertebrae  [C0024091] [28] & 0.77 & 0.77 & 0.76\\
│\verb!   !│\verb!   !└── paravertebral  [C0442150] [41] & 0.65 & 0.66 & 0.61\\
│\verb!   !└── rib  [C0035561] [760] & 0.72 & 0.66 & 0.68\\
│\verb!   !\verb!   ! ├── anterior rib  [C4323264] [73] & 0.66 & 0.60 & 0.61\\
│\verb!   !\verb!   ! ├── posterior rib  [C4323265] [134] & 0.68 & 0.67 & 0.69\\
│\verb!   !\verb!   ! └── rib cartilage  [C0222787] [7] & 0.75 & 0.77 & 0.74\\
├── hemithorax  [C0934569] [749] & 0.84 & 0.81 & 0.83\\
├── extrapleural  [C0442091] [21] & 0.78 & 0.71 & 0.75\\
├── extrapulmonary   [9] & 0.68 & 0.64 & 0.64\\
├── pleural  [C0032225] [2156] & 0.84 & 0.80 & 0.84\\
├── subpleural  [C0225775] [40] & 0.89 & 0.84 & 0.86\\
├── fissure  [C0458078] [147] & 0.85 & 0.75 & 0.82\\
│\verb!   !├── minor fissure  [C0734040] [83] & 0.91 & 0.77 & 0.86\\
│\verb!   !└── major fissure  [C4253583] [28] & 0.81 & 0.75 & 0.79\\
├── lobar  [C0225752] [604] & 0.79 & 0.76 & 0.78\\
│\verb!   !├── upper lobe  [C0225756] [1838] & 0.86 & 0.78 & 0.84\\
│\verb!   !│\verb!   !├── left upper lobe  [C1261076] [298] & 0.83 & 0.76 & 0.81\\
│\verb!   !│\verb!   !│\verb!   !└── lingula  [C0225740] [214] & 0.82 & 0.69 & 0.77\\
│\verb!   !│\verb!   !└── right upper lobe  [C1261074] [568] & 0.84 & 0.76 & 0.83\\
│\verb!   !├── lower lobe  [C0225758] [774] & 0.82 & 0.77 & 0.81\\
│\verb!   !│\verb!   !├── left lower lobe  [C1261077] [348] & 0.81 & 0.75 & 0.80\\
│\verb!   !│\verb!   !└── right lower lobe  [C1261075] [228] & 0.81 & 0.75 & 0.79\\
│\verb!   !└── middle lobe  [C4281590] [328] & 0.80 & 0.74 & 0.77\\
├── subsegmental  [C0929165] [351] & 0.81 & 0.69 & 0.75\\
├── bronchi  [C0006255] [621] & 0.77 & 0.75 & 0.76\\
├── peribronchi  [C0225607] [24] & 0.85 & 0.76 & 0.82\\
├── lung field  [C0225759] [7703] & 0.83 & 0.78 & 0.82\\
│\verb!   !├── pleural  [C0032225] [2156] & 0.84 & 0.80 & 0.84\\
│\verb!   !├── subpleural  [C0225775] [40] & 0.89 & 0.84 & 0.86\\
│\verb!   !├── major fissure  [C4253583] [28] & 0.81 & 0.75 & 0.79\\
│\verb!   !├── subsegmental  [C0929165] [351] & 0.81 & 0.69 & 0.75\\
│\verb!   !├── bronchi  [C0006255] [621] & 0.77 & 0.75 & 0.76\\
│\verb!   !├── peribronchi  [C0225607] [24] & 0.85 & 0.76 & 0.82\\
│\verb!   !├── upper lung field  [C0929227] [1944] & 0.86 & 0.78 & 0.84\\
│\verb!   !│\verb!   !└── upper lobe  [C0225756] [1838] & 0.86 & 0.78 & 0.84\\
│\verb!   !│\verb!   !\verb!   ! ├── left upper lobe  [C1261076] [298] & 0.83 & 0.76 & 0.81\\
│\verb!   !│\verb!   !\verb!   ! ├── right upper lobe  [C1261074] [568] & 0.84 & 0.76 & 0.83\\
│\verb!   !│\verb!   !\verb!   ! ├── apical  [C0734296] [872] & 0.86 & 0.80 & 0.84\\
│\verb!   !│\verb!   !\verb!   ! └── suprahilar  [C0934260] [49] & 0.64 & 0.60 & 0.63\\
│\verb!   !├── middle lung field  [C0929434] [1960] & 0.76 & 0.73 & 0.75\\
│\verb!   !│\verb!   !├── aortopulmonary window  [C1282038] [32] & 0.61 & 0.62 & 0.60\\
│\verb!   !│\verb!   !├── hilar  [C0205150] [1533] & 0.75 & 0.71 & 0.73\\
│\verb!   !│\verb!   !│\verb!   !├── pulmonary artery  [C0034052] [38] & 0.69 & 0.64 & 0.65\\
│\verb!   !│\verb!   !│\verb!   !├── hilar bilateral   [86] & 0.82 & 0.76 & 0.79\\
│\verb!   !│\verb!   !│\verb!   !└── perihilar  [C0225702] [335] & 0.85 & 0.81 & 0.82\\
│\verb!   !│\verb!   !└── minor fissure  [C0734040] [83] & 0.91 & 0.77 & 0.86\\
│\verb!   !└── lower lung field   [4559] & 0.81 & 0.76 & 0.80\\
│\verb!   !\verb!   ! ├── basal  [C1282378] [1625] & 0.80 & 0.75 & 0.78\\
│\verb!   !\verb!   ! ├── lower lobe  [C0225758] [774] & 0.82 & 0.77 & 0.81\\
│\verb!   !\verb!   ! │\verb!   !├── left lower lobe  [C1261077] [348] & 0.81 & 0.75 & 0.80\\
│\verb!   !\verb!   ! │\verb!   !└── right lower lobe  [C1261075] [228] & 0.81 & 0.75 & 0.79\\
│\verb!   !\verb!   ! ├── middle lobe  [C4281590] [328] & 0.80 & 0.74 & 0.77\\
│\verb!   !\verb!   ! ├── infrahilar   [108] & 0.77 & 0.70 & 0.77\\
│\verb!   !\verb!   ! ├── lingula  [C0225740] [214] & 0.82 & 0.69 & 0.77\\
│\verb!   !\verb!   ! ├── supradiaphragm   [2] & 0.84 & 0.48 & 0.69\\
│\verb!   !\verb!   ! ├── diaphragm  [C0011980] [721] & 0.76 & 0.69 & 0.73\\
│\verb!   !\verb!   ! ├── infradiaphragm   [25] & 0.84 & 0.77 & 0.81\\
│\verb!   !\verb!   ! ├── cardiophrenic angle   [67] & 0.66 & 0.66 & 0.64\\
│\verb!   !\verb!   ! └── costophrenic angle  [C0230151] [1235] & 0.79 & 0.72 & 0.77\\
│\verb!   !\verb!   !\verb!   !  ├── right costophrenic angle  [C0504099] [285] & 0.85 & 0.77 & 0.81\\
│\verb!   !\verb!   !\verb!   !  ├── left costophrenic angle  [C0504100] [359] & 0.86 & 0.71 & 0.82\\
│\verb!   !\verb!   !\verb!   !  └── bilateral costophrenic angle   [10] & 0.87 & 0.85 & 0.91\\
├── peripheral  [C0205100] [241] & 0.95 & 0.93 & 0.93\\
├── central  [C0205099] [111] & 0.84 & 0.83 & 0.83\\
├── mediastinum  [C0025066] [5136] & 0.80 & 0.78 & 0.79\\
│\verb!   !├── superior mediastinum  [C0230147] [538] & 0.84 & 0.80 & 0.83\\
│\verb!   !│\verb!   !├── carotid artery  [C0007272] [1] & 1.00 & 0.55 & 1.00\\
│\verb!   !│\verb!   !├── brachiocephalic veins  [C0006095] [7] & 0.86 & 0.82 & 0.88\\
│\verb!   !│\verb!   !├── supra aortic   [228] & 0.87 & 0.84 & 0.85\\
│\verb!   !│\verb!   !├── aortic button  [C0003489] [80] & 0.78 & 0.75 & 0.76\\
│\verb!   !│\verb!   !├── superior cave vein  [C3165182] [77] & 0.94 & 0.90 & 0.94\\
│\verb!   !│\verb!   !└── subclavian vein  [C0038532] [117] & 0.96 & 0.94 & 0.96\\
│\verb!   !├── lower mediastinum  [C1261193] [3489] & 0.78 & 0.76 & 0.77\\
│\verb!   !│\verb!   !├── anterior mediastinum  [C0230148] [17] & 0.63 & 0.69 & 0.66\\
│\verb!   !│\verb!   !│\verb!   !└── thymus  [C0040113] [7] & 0.61 & 0.74 & 0.53\\
│\verb!   !│\verb!   !├── middle mediastinum  [C0230149] [3142] & 0.80 & 0.78 & 0.79\\
│\verb!   !│\verb!   !│\verb!   !└── cardiac  [C1522601] [3134] & 0.80 & 0.78 & 0.79\\
│\verb!   !│\verb!   !│\verb!   !\verb!   ! └── coronary  [C1522318] [12] & 0.82 & 0.79 & 0.82\\
│\verb!   !│\verb!   !└── posterior mediastinum  [C0230150] [363] & 0.73 & 0.69 & 0.71\\
│\verb!   !│\verb!   !\verb!   ! └── retrocardiac   [361] & 0.73 & 0.69 & 0.72\\
│\verb!   !└── aortic  [C0003483] [2059] & 0.88 & 0.85 & 0.87\\
├── paratracheal  [C0442143] [91] & 0.65 & 0.64 & 0.68\\
├── airways  [C0458827] [900] & 0.73 & 0.71 & 0.72\\
│\verb!   !├── tracheal  [C0040578] [294] & 0.79 & 0.66 & 0.80\\
│\verb!   !└── bronchi  [C0006255] [621] & 0.77 & 0.75 & 0.76\\
├── esophageal  [C1522619] [24] & 0.68 & 0.66 & 0.65\\
├── paramediastinum   [19] & 0.80 & 0.79 & 0.81\\
├── paracardiac   [40] & 0.76 & 0.69 & 0.72\\
├── epigastric   [10] & 0.69 & 0.62 & 0.60\\
├── gastric chamber  [C3714551] [20] & 0.81 & 0.74 & 0.71\\
├── hypochondrium  [C0230186] [40] & 0.66 & 0.68 & 0.71\\
│\verb!   !├── right hypochondrium  [C0738590] [25] & 0.63 & 0.62 & 0.69\\
│\verb!   !│\verb!   !└── gallbladder  [C0016976] [13] & 0.63 & 0.63 & 0.67\\
│\verb!   !└── left hypochondrium  [C0738591] [15] & 0.68 & 0.73 & 0.76\\
├── right  [C0444532] [3594] & 0.76 & 0.72 & 0.74\\
├── left  [C0443246] [2877] & 0.77 & 0.73 & 0.76\\
└── bilateral  [C0238767] [1913] & 0.82 & 0.80 & 0.80\\
\verb!   ! ├── diffuse bilateral   [149] & 0.91 & 0.86 & 0.87\\
\verb!   ! └── basal bilateral   [435] & 0.82 & 0.77 & 0.79\\
\hline

\end{longtable}


\end{document}